\documentclass[aps,prl,twocolumn,showpacs,amsmath,preprintnumbers]{revtex4}
\usepackage{hyperref}
\begin{document}

\preprint{} 
\preprint{gr-qc/0702107}
\title{Black Hole Thermodynamics from Euclidean Horizon Constraints}
\author{S.\ Carlip}
\email[]{carlip@physics.ucdavis.edu}
\affiliation{Department of Physics,
University of California, Davis, CA 95616}

\date{\today}

\begin{abstract}
To explain black hole thermodynamics in quantum gravity, one must 
introduce constraints to ensure that a black hole is actually present.  
I show that for a large class of black holes, such ``horizon constraints'' 
allow the use of conformal field theory techniques to compute the 
density of states, reproducing the Bekenstein-Hawking entropy in a 
nearly model-independent manner.  One standard string theory approach 
to black hole entropy arises as a special case, lending support to the 
claim that the mechanism may be ``universal.''  I argue that the relevant 
degrees of freedom are Goldstone-boson-like excitations arising from the 
weak breaking of symmetry by the constraints.
\end{abstract}

 
\maketitle
Thirty years ago, Bekenstein and Hawking showed us that black holes are 
thermodynamic systems, with characteristic temperatures and entropies 
\cite{Hawking}.  This poses a problem: in Wheeler's words, black holes 
have ``no hair,'' no classical degrees of freedom that could account for 
such thermal behavior.  The most likely resolution is that the relevant 
degrees of freedom are purely quantum mechanical, in which case black hole 
statistical mechanics might offer important insights into quantum 
gravity.

Until recently, little was known about such quantum degrees of freedom.  Today,
we suffer an embarrassment of riches.  The microscopic states of a black hole 
can be counted in at least three ways, not obviously equivalent, in string 
theory \cite{StromVafa,Skenderis,Mathur}; in two rather different approaches 
to loop quantum gravity \cite{Ash,LivTerno}; in induced gravity \cite{Frolov}; 
and perhaps in causal set models \cite{Rideout}.  Each of these approaches
has limitations, but each seems to work within its domain of applicability.  The 
new puzzle is that despite counting very different degrees of freedom, they
all give the same answer.  More generally, in the absence of any classical 
degrees of freedom to which one could apply the correspondence principle,
it is not obvious why \emph{any} counting of microscopic states should 
reproduce Hawking's original semiclassical results.

One possible answer, first suggested in \cite{Carlipa}, is that a classical
symmetry near the black hole horizon might control the number of degrees of
freedom, independent of the details of quantum gravity.  To an observer
outside a black hole, the near-horizon region is effectively two-dimensional
and conformally invariant \cite{Birmingham}: transverse excitations and 
dimensionful quantities are red-shifted away relative to the degrees of freedom 
in the $r$-$t$ plane.  Such a conformal description is powerful enough to 
determine the flux and spectrum of Hawking radiation \cite{Wilczek}. 
But as Cardy has shown \cite{Cardy}, the density of states, and thus the
entropy, of a two-dimensional conformal field theory are completely fixed 
by a small number of parameters, independent of the particulars of the theory.

In this letter, I compute these parameters for a very large class of black 
holes and obtain the usual Bekenstein-Hawking entropy.  The key step is the 
imposition of ``horizon constraints'' to ensure the 
presence of a black hole.  I show that one standard string theory approach 
occurs as a special case, offering support for the claim of ``universality.''  
Finally, I argue that conformal methods suggest a physical picture of 
  the relevant degrees of freedom, as Goldstone-boson-like 
excitations arising from the breaking of symmetry by the conformal anomaly.

\section{Euclidean dilaton gravity}
If the relevant near-horizon degrees of freedom are two-dimensional, 
a two-dimensional  effective field theory description should exist.  For 
the case of spherical symmetry, the Kaluza-Klein reduction to two dimensions 
is well understood.  It is less well known that a similar reduction exists
even without symmetries \cite{Yoon}, albeit at the expense of introducing 
an enormous gauge group.   

Given such an effective theory, we must next ensure that a black hole 
is actually present, by imposing boundary conditions or constraints that imply 
the existence of a horizon.  The ``horizon as boundary'' approach of \cite{Carlipa} 
has had some success, but involves transformations that diverge at the horizon 
\cite{Dreyer}.  A direct imposition of ``horizon constraints'' was attempted in 
\cite{Carlipc}, but ambiguities remain, traceable to the lightlike nature of 
the horizon. 

To avoid these difficulties, I will follow the common path of analytically 
continuing to ``Euclidean'' gravity.  This shrinks the horizon from a light cone 
$t^2-r_*^2=0$ to a point $t^2 + r_*^2=0$, and makes time periodic. As in string 
theory, I will consider ```radial quantization,'' evolution outward in a radial 
direction.  After a field redefinition, the two-dimensional dilaton gravity action 
is then \cite{MartKun}
\begin{align}
I = \frac{1}{2}\int d^2x\sqrt{g}\left[ \varphi R 
  + V[\varphi] - \frac{1}{2} W[\varphi]h_{IJ}F^I_{ab}F^{Jab}\right] 
\label{a1} 
\end{align}
where $F^I_{ab}$ is a non-Abelian gauge field strength, $V[\varphi]$ 
and $W[\varphi]$ are arbitrary potentials, and my units are $8\pi G=1$.  For 
dimensionally reduced gravity, the dilaton field $\varphi$ is the transverse area; 
in general, the entropy of a black hole is proportional to its value $\varphi_H$ 
at the horizon.

Introducing an ADM-like decomposition of the metric 
\begin{align}
ds^2 = N^2f^2dr^2 + f^2(dt + \alpha dr)^2 
\label{a2}
\end{align}
and transforming to (radial) Hamiltonian form, one finds that the Hamiltonian
is, as usual, a sum of constraints,
\begin{align}
&{\cal H}_\parallel = {\dot\varphi}\pi_\varphi - f{\dot\pi}_f = 0\nonumber\\
&{\cal H}_\perp = f\pi_f\pi_\varphi + f\left(\frac{\dot\varphi}{f}\right)^\cdot
  -\frac{1}{2}f^2{\hat V} = 0 \label{a4}\\
&{\cal H}_I = {\dot\pi}_I - c^J{}_{IK}A^K\pi_J = 0 \nonumber 
\end{align}
with $A^I = A_t^I$ and ${\hat V} = V + h^{IJ}\pi_I\pi_J/W$.  By the ${\cal H}_I$ 
constraints, $h^{IJ}\pi_I\pi_J$ is a constant, so the net effect of the gauge 
field is simply to shift $V$.  The constraints ${\cal H}_\parallel$ and 
${\cal H}_\perp$ can be combined into two complex generators,
\begin{align}
L[\xi] &= \frac{1}{2}\int dt\xi({\cal H}_\parallel + i{\cal H}_\perp) \nonumber\\
{\bar L}[\xi] &= \frac{1}{2}\int dt\xi({\cal H}_\parallel - i{\cal H}_\perp) .
\label{a5}
\end{align}
These generate holomorphic and antiholomorphic diffeomorphisms, the fundamental
symmetries of a conformal field theory, and satisfy the 
Virasoro algebra \cite{CFT}
\begin{align}
\left\{L[\xi],L[\eta]\right\} &= L[\xi{\dot\eta}-\eta{\dot\xi}]
  + \frac{c}{48\pi}\int dt ({\dot\xi}{\ddot\eta}-{\dot\eta}{\ddot\xi})
  \nonumber \\
\left\{L[\xi],{\bar L}[\eta]\right\} &= 0\label{a6}\\
\left\{{\bar L}[\xi],{\bar L}[\eta]\right\} 
  &= {\bar L}[\xi{\dot\eta}-\eta{\dot\xi}]
  + \frac{{\bar c}}{48\pi}\int dt ({\dot\xi}{\ddot\eta}-{\dot\eta}{\ddot\xi})
  \nonumber
\end{align}
with central charges $c={\bar c} = 0$.

\section{Horizon constraints}
We next demand that a point $r=r_H$ be a black hole horizon.  The 
necessary conditions were worked out by Teitelboim \cite{Teitelboim}: 
we require $s={\bar s}=0$ at $r_H$, where
\begin{align}
s = f\pi_f - i{\dot\varphi} .
\label{b1}
\end{align}
Unfortunately, our radial foliation breaks down at $r_H$, so we must be
slightly indirect, requiring a ``stretched horizon'' at $r_H+\epsilon$ 
and then take the limit $\epsilon\rightarrow0$.

To find the appropriate stretched horizon conditions, note that a black hole 
is characterized by two quantities, the expansion $\vartheta$ and the surface 
gravity $\hat\kappa$.  The expansion, the fractional rate of change of the 
transverse area, is simply $\vartheta=s/\varphi$.
The surface gravity is trickier: it depends on the normalization of the null
normal to the horizon, which in turn depends on asymptotic behavior 
\cite{Ashb}.  We can initially define $\kappa = \pi_\varphi - i{\dot f}/{f}$.  
Then under a rescaling of the null normal, or equivalently a conformal 
transformation $f^2\rightarrow e^{2\omega}f^2$ (with the conformal factor 
$\omega$ to be determined later), we obtain 
\begin{align}
{\hat\kappa} = \kappa + f^2\frac{d\omega}{d\varphi} 
 = \pi_\varphi - i{\dot f}/{f} + f^2\frac{d\omega}{d\varphi}.
\label{b3}
\end{align}

As our horizon constraints, we now choose
\begin{align}
K &= s - a({\hat\kappa} - {\hat\kappa}_H) = 0 \nonumber\\
{\bar K} &= {\bar s} - a({\bar{\hat\kappa}} - {\bar{\hat\kappa}}_H) = 0
\label{b4}
\end{align}
where the constant $a$ will be determined below.  The $\hat\kappa$ dependence 
is fixed by requiring the existence of a Hamiltonian for conformal transformations 
at the horizon \cite{Carlipd}, and ensures invariance under constant rescalings 
of the horizon normal \cite{Carlipc}.  Our constraints define a stretched 
horizon around a black hole with surface gravity ${\hat\kappa}_H$, and reduce 
to Teitelboim's conditions as ${\hat\kappa}\rightarrow{\hat\kappa}_H$.

\section{Horizon algebra and state-counting}
We now make the crucial observation that the horizon constraints (\ref{b4}) are
not preserved by the symmetries (\ref{a5}).  We can cure this  
with a trick introduced by Bergmann and Komar \cite{Bergmann}: we add ``zero,'' 
in the form of  $K$ and $\bar K$, to the Virasoro generators to produce new 
generators that commute with $K$ and $\bar K$.  Specifically, if $\Delta_{ij}$ 
is the inverse of $\{K_i,K_j\}$, then any modified observable
\begin{align}
O^* = O - \sum_{i,j}\int dudv\{O,K_i(u)\}\Delta_{ij}(u,v)K_j(v)
\label{c2}
\end{align}
will have vanishing brackets with the $K_i$.  

The Poisson brackets of such modified observables are equivalent to Dirac brackets 
with $K$ and $\bar K$ treated as second class constraints.  A direct calculation 
yields
\begin{align}
\left\{L^*[\xi],L^*[\eta]\right\} =& L^*[\xi{\dot\eta}-\eta{\dot\xi}]\nonumber\\ 
  &-\frac{ia}{8}\int dt ({\dot\xi}{\ddot\eta}-{\dot\eta}{\ddot\xi}) 
  + \frac{i}{8}\int dt A(\xi{\dot\eta}-\eta{\dot\xi}) 
  \nonumber\\
\left\{L^*[\xi],{\bar L}^*[\eta]\right\} =& \int dt 
  [B_1(\xi{\dot\eta}-\eta{\dot\xi}) + B_2\xi\eta]
  \label{c3}\\
\left\{{\bar L}^*[\xi],{\bar L}^*[\eta]\right\} =& {\bar L}^*[\xi{\dot\eta}-\eta{\dot\xi}]
  \nonumber\\ 
  &+ \frac{ia}{8}\int dt ({\dot\xi}{\ddot\eta}-{\dot\eta}{\ddot\xi}) 
  - \frac{i}{8}\int dt {\bar A}(\xi{\dot\eta}-\eta{\dot\xi})
  \nonumber
\end{align}
with
\begin{align}
&A = a{\hat\kappa}_H^2 
   + f^2\left({\hat V} - \frac{a}{2}\frac{d{\hat V}}{d\varphi} 
   - 2a{\bar{\hat\kappa}}_H\frac{d\omega}{d\varphi}\right) + {\cal O}(\varphi-\varphi_H)^2
   \nonumber\\
&B_1 , B_2 = {\cal O}(\varphi-\varphi_H)^2  . \label{c4}
\end{align}

If we now demand that the ${\cal O}(\varphi-\varphi_H)$ term in the anomaly $A$ vanish,  
the conformal factor $\omega$ is fixed:
\begin{align}
\omega = -\frac{{\hat V}_H}{4{\hat{\bar\kappa}}_H}\left[\ln(\varphi{\hat V})
  - \left(1+\frac{2\varphi_H}{a}\right)\ln\varphi\right]
  + {\cal O}(\varphi-\varphi_H)^2.
\label{c4a}
\end{align}
With this choice, (\ref{c3}) is again a Virasoro algebra at the horizon, but with 
a shifted central charge and zero-modes.  Choosing a basis of periodic 
vector fields
\begin{align}
\xi_n = \frac{\beta}{2\pi}e^{2\pi int/\beta} ,
\label{c5}
\end{align}
we can read off central charges and conformal weights $\Delta=L[\xi_0]$,
${\bar\Delta}={\bar L}[\xi_0]$: with factors of $8\pi G$ restored,
\begin{align}
c = {\bar c} = \frac{3a}{4G} ,\qquad 
\Delta = {\bar\Delta} = \frac{a}{32G}\left(\frac{\kappa_H\beta}{2\pi}\right)^2 .
\label{c6}
\end{align}
Cardy's formula \cite{Cardy} then determines the entropy:
\begin{align}
S = 2\pi\sqrt{\frac{c\Delta}{6}} + 2\pi\sqrt{\frac{{\bar c}{\bar \Delta}}{6}}
  = \frac{\pi|a|}{4G}\left(\frac{\kappa_H\beta}{2\pi}\right) .
\label{c7}
\end{align}

We must still fix the constants $a$ and $\beta$.  We begin by finding the classical 
solutions.  Defining a new function ${\hat\jmath}[\varphi]$ by the condition 
${\hat V}[\varphi] = {d{\hat\jmath}}/{d\varphi}$, it is straightforward to show 
that the general static solution is
\begin{align} 
f^2 = s = {\hat\jmath} - {\hat\jmath}_H, \ \varphi' = N({\hat\jmath} - {\hat\jmath}_H) ,  
\ \kappa = {\hat V}/2 ,
\label{d2}
\end{align}
where $N$ is arbitrary, ${\hat\jmath}_H$ is an integration constant, and a horizon occurs
at $\varphi_H={\hat\jmath}^{-1}({\hat\jmath}_H)$.  The metric (\ref{a2}) has a conical 
singularity at the horizon unless $t$ has a period
\begin{align}
\beta = {2\pi}/{{\hat\kappa}_H} ,
\label{d3}
\end{align}
giving the usual Hawking temperature.  With exceptions I will describe in the next
section, we should thus take this to be the period appearing in (\ref{c5}) and 
(\ref{c7}).

As noted earlier, the surface gravity $\hat\kappa$ still has an ambiguity, which 
we must fix by looking at asymptotic behavior.  It was observed in \cite{Kunstatter} 
that the the metric can be put into asymptotically Schwarzschild form by a conformal 
transformation $e^{2\omega}=1/{\hat\jmath}$.  We cannot use this here, since we 
know $\hat\jmath$ only up to an additive constant determined by a particular 
solution; we do not want our definition of $\hat\kappa$ to depend on what solution 
we are looking at.  But for any solution,
$f^2 = {\hat V}\varphi(1 - \varphi_H/\varphi) + {\cal O}(1-\varphi_H/\varphi)^2$   
near the horizon.  We can therefore take $e^{2\omega} = 1/{\hat V}\varphi$.
This precisely matches the condition (\ref{c4a}) for anomaly cancellation,
and further fixes $a=-2\varphi_H$. 

The entropy (\ref{c7}) is thus $S=2\pi\varphi_H/4G$.  This is \emph{almost} the 
standard Bekenstein-Hawking entropy; it differs by a factor of $2\pi$.  The same 
factor appears in \cite{Park}, and has a direct physical explanation: black hole 
entropy counts horizon degrees of freedom at a fixed time, while we have computed 
the entropy at the horizon for \emph{all} times, effectively integrating 
over a circle of circumference $2\pi$.

Up to this $2\pi$ factor, the central charge (\ref{c6}) agrees with the ``horizon as
boundary'' results of \cite{Carlipa}, and the conformal weights match a two-dimensional
nonchiral version of the Komar integral of \cite{EmpMat}.  The relationship to loop 
quantization is less clear, but the loop horizon states in \cite{Ash} trace back to 
an $\hbox{SL}(2,{\bf R})$ Chern-Simons theory with coupling constant $k=iA/8\pi\gamma G$.  
Such a Chern-Simons theory induces a Liouville theory at boundaries with central charge 
$c=6k$ \cite{Carlipe}; for the natural value $\gamma=i$ of the Immirzi parameter, this 
again matches our results.

Our central charge does not, however, match that of the BTZ black hole.  Since the 
BTZ entropy calculation is a basic element of one string theory approach to black hole 
entropy, let us examine this case more closely.

\section{BTZ and stringy black holes}

The BTZ black hole \cite{Carlipe} is a rotating black hole in (2+1)-dimensional 
gravity with cosmological constant $\Lambda=-1/\ell^2$.  It has an event horizon at 
$r_+$, an inner horizon at $r_-$, and surface gravity $\kappa_{\hbox{\tiny\it BTZ}} %
= {(r_+^2-r_-^2)}/{\ell^2r_+}$.  It is asymptotically anti-de Sitter, with a boundary 
at infinity that is geometrically a flat cylinder; the asymptotic symmetry group is 
described by a Virasoro algebra with central charge and conformal weight
\begin{align}
c = {\bar c} = \frac{3\ell}{2G}, \quad \Delta = \frac{(r_++r_-)^2}{16G\ell}, \ 
  {\bar\Delta} = \frac{(r_+-r_-)^2}{16G\ell} .
\label{e2}
\end{align}
With these parameters, the Cardy formula (\ref{c7}) gives the standard Bekenstein-Hawking 
entropy of $2\pi r_+/4G$.

This result depends on symmetries at infinity, and thus differs from the approach 
here.  But (2+1)-dimensional gravity has no local degrees of freedom, so we should 
be able to translate the BTZ results to the horizon.  To do so, note first that 
the vectors generating the symmetries of the BTZ black hole differ from (\ref{c5}):
instead,
\begin{align}
\xi_n^{\hbox{\tiny\it BTZ}} \sim e^{in(t-\ell\phi)/\ell} , \qquad 
{\bar\xi}_n^{\hbox{\tiny\it BTZ}} \sim e^{in(t+\ell\phi)/\ell} ,
\label{e3}
\end{align}
where $\phi$ is an angular coordinate.  For the nonrotating case, 
we can change the period from $2\pi/\kappa_{\hbox{\tiny\it BTZ}}=2\pi\ell^2/r_+$ 
(as in(\ref{c5})) to $2\pi\ell$ (as in (\ref{e3})) by restricting to modes $Nn$,
with $N=\ell/r_+$.  But it is known that such a restriction results in
a new Virasoro algebra, with central charge and conformal weight ${\tilde c}=cN$, 
${\tilde\Delta}=\Delta/N$ \cite{Banados}.  For our $N$, the new central charges 
${\tilde c} = {\tilde{\bar c}} = ({3a}/{4G})\cdot({\ell}/{r_+})$ then agree with 
the BTZ values (\ref{e2}), up to the factor of $2\pi$ discussed earlier.

The shifted conformal weights are now ${\tilde\Delta} = {\tilde{\bar\Delta}} %
= 2\pi(r_+^2/16G\ell)(\kappa\beta/2\pi)$.  These match the BTZ values when 
$r_-=0$, but differ for spinning black holes.  To understand this difference, 
note that the angular coordinate $\phi$ in (\ref{e3}) is that of an observer at 
infinity.  For dimensional reduction, we should instead use a corotating coordinate 
at the horizon, $\phi' = \phi - (r_-/r_+\ell)t$.  This alters the time dependence 
of the modes (\ref{e3}), shifting their periods to $\beta_\pm = (1\pm r_-/r_+)\beta$.  
These new periods give the ``left'' and ``right'' Hawking temperatures that appear 
in string theory, and, inserted into (\ref{c6}), reproduce the BTZ weights (\ref{e2}),
again up to a factor of $2\pi$.

The conformal computation of the BTZ black hole entropy can thus be viewed as a
special case of our horizon constraint method, suitably translated to spatial
infinity.  By itself, this might not be of great significance.  But a standard 
approach to entropy in string theory \cite{Skenderis} uses the BTZ-like near-horizon 
structure of near-extremal black holes, combined with various duality transformations, 
to reduce the problem to that of the BTZ black hole.  The methods of this letter 
thus encompasses an important string theory approach to black hole entropy.

\section{What are we counting?}

A key advantage of the horizon constraint approach is that while it lets us counts 
states, it does not require any detailed knowledge of the states being counted.  
Nevertheless, we can interpret these results as suggesting an interesting effective 
description of the states responsible for black hole entropy.
In conventional approaches to quantum gravity, physical states are required to be 
invariant under diffeomorphisms; that is, in our context,
\begin{align}
L[\xi]|\mathit{phys}\rangle = {\bar L}[\xi]|\mathit{phys}\rangle = 0.
\label{f1}
\end{align}
If the Virasoro algebra (\ref{a6}) has a central charge, however, such conditions
are incompatible with the Poisson brackets.  We know how to weaken (\ref{f1}):
we can require, for example, that only positive frequency modes annihilate physical 
states \cite{CFT}.  But then new states---for instance, the descendant states 
$L_{-n}|0\rangle$---become physical.

This phenomenon is reminiscent of the Goldstone mechanism \cite{KalTern}: the 
conformal anomaly breaks diffeomorphism invariance, and ``would-be pure gauge'' 
degrees of freedom become physical.  For asymptotically anti-de Sitter spacetimes
in three and five dimensions, an explicit description of the corresponding degrees 
of freedom at infinity is possible \cite{Carlipf}; one might hope for something
similar at a horizon.

Perhaps the most important open question is whether we can also incorporate Hawking 
radiation and black hole evaporation.  As noted above, one can compute the Hawking 
radiation spectrum from methods based on conformal anomalies of matter fields
\cite{Wilczek}.  In \cite{Emparan}, it was shown that a scalar field could be 
coupled to the conformal boundary degrees of freedom of the BTZ black hole to
obtain Hawking radiation; if a similar coupling were possible here, it would 
represent major progress.

\section{Acknowledgments}
This work was supported in part by U.S.\ 
Department of Energy grant DE-FG03-91ER40674.

\end{document}